\documentclass{elsarticle}
\usepackage{amsmath}
\journal{Physica B: Condensed Matter}

\begin{document}

\begin{frontmatter}

\title{Novel $\pi$-type vortex in a nanoscale extreme type-II superconductor: Induced by quantum-size effect}

\author{Haiyan Huang}
\author{Qing Liu}
\author{Wenhui Zhang}
\author{Yajiang Chen\corref{mycorrespondingauthor}}
\address{College of Engineering and Design, Lishui University, Zhejiang 323000, China}

\cortext[mycorrespondingauthor]{Email address: Yajiang.Chen@lsu.edu.cn}
%\ead{Yajiang.Chen@lsu.edu.cn}

\begin{abstract}
By numerically solving the Bogoliubov-de Gennes equations, we report a novel $\pi$-type vortex state whose order parameter near the core undergoes an extraordinary $\pi$-phase change for a quantum-confined extreme type-II $s$-wave superconductor. Its supercurrent behaves as the cube of the radial coordinate near the core, and its local density of states spectrum exhibits a significant negative-shifted zero-bias peak. Such $\pi$-type vortex state is induced by quantum-size effect, and can survive thermal smearing at temperatures up to a critical value $T_\tau$. The Anderson's approximation indicates that the $\pi$-type vortex may remain stable under sufficiently week magnetic field in the case less deep in the type-II limit. Moreover, we find that its appearance is governed by the sample size and $k_F\xi_0$ with $k_F$ the Fermi wave number and $\xi_0$ the zero-temperature coherence length. Similar effects may be expected in quantum-confined ultracold superfluid Fermi gasses, or even high-$T_c$ superconductors with proper $k_F\xi_0$ value.
\end{abstract}


\end{frontmatter}

%\linenumbers

\section{Introduction}\label{sec:intro}
Recently, $\pi$-type Meissner state, in which the spatial order
parameter changes sign, has been found in many circumstances, e.g. at a superconductor-ferromagnet interface~\cite{Ryazanov2001,Krivoruchko2002, Buzdin2005}, in $d$-wave superconductors~\cite{VanHarlingen1995}, and near an impurity site in $s$-wave superconductors~\cite{Balatsky2006}. From these cases, it seams that the existence of $\pi$-type state in superconductors requires two necessary factors: inhomogeneous superconductivity and interactions suppressing superconductivity. Superconducting vortex satisfies these conditions: a normal-superconducting transition domain is created by magnetic flux around the core and the system shows inhomogeneous superconductivity, $\pi$-phase shift in vortex states has not been discovered yet.

Caroli, de Gennes and Matricon~\cite{Caroli1964} first discovered the microstructure of a superconducting vortex strongly depends on in-gap fermions, and the corresponding pronounced symmetric zero bias peak (ZBP) of the local density of states (LDOS) at the core was confirmed in experiments~\cite{Hess1989,Hess1990}. Recently, it was found~\cite{Roditchev2015} that the quantum interference of Andreev quasiparticles can induce Josephson vortex. For a macroscopic single-quantum vortex state, the energies and angular momenta of these in-gap fermions have the same sign, such chirality leads to a collapse of the vortex core in a clean $s$-wave superconductor near zero temperature ($T$). It is called the Kramer-Pesch (KP) effect~\cite{Kramer1974}. Similar behavior was also found in bulk superfluid fermionic condensates~\cite{Volovik1993}, e.g. ultracold Fermi gases~\cite{Nygaard2003} and superfluid neutron star matter~\cite{DeBlasio1999}.

Scenario may change dramatically in the regime of nanoscale superconductivity, where quantum confinement breaks the translational symmetry, and strongly modulates the behavior of quasiparticles. It has been reported that the quantum confinement leads to inhomogeneous superconductivity even for the Meissner state~\cite{Guo2004,Shanenko2007,Chen2009,Chen2010}. The existence of single- and multi-quanta vortex states has already been verified in high-quality superconducting Pb nanofilms by the scanning tunnelling spectroscopy~\cite{Cren2009,Cren2011}. The calculations~\cite{Zhang2012, Zhang2013} based on the Bogoliubov-de Gennes theory also show that the quantum confinement may cause several unconventional vortex states, such as asymmetric ones and giant-multivortex combinations. For an isolated vortex in quantum-confined systems, we previously found~\cite{Chen2014,Chen2015} that quantum-size effect may break down the chirality of in-gap fermions with small angular momenta, and cause a significant extension of the vortex core at finite temperatures, i.e. the superconductivity near the core is surprisedly depressed. Therefore, a natural question arises: whether such depression can produce an extraordinary vortex state with a $\pi$-phase shift in the order parameter near to the core.

In this paper we numerically solve the Bogoliubov-de Gennes (BdG) equations for a single-quantum vortex in a nanoscale cylindrical extreme type-II superconductor, and find that a $\pi$-phase shift may occur in the radial order parameter near the core at $T=0$ K when the chirality of the in-gap fermion with the lowest azimuthal quantum number $m$ breaks down due to quantum confinement. Such $\pi$-type vortex state shows $\rho^3$-dependence in supercurrent $j(\rho)$. Moreover, we find that $k_{F}\xi_0$ and radius $R$ of the sample control the chirality breakdown of the in-gap fermion with the lowest $m$, i.e. the appearance of the $\pi$-type vortex state. Similar behavior may also be expected in other nanoscale extreme type-II superconductors, e.g. metallic nanoislands on Si~\cite{Cren2009,Cren2011}, and ultracold superfluid Fermi gases in cigar- and pancake-shaped atomic traps~\cite{Bloch2008}.

The paper is organized as follows. In Sec.~\ref{theory}, we outline the theoretical model based on the BdG equations for a single-quantum vortex in an extreme type-II $s$-wave superconducting nanocylinder. In Sec.~\ref{results}, the behavior of the radial order parameter and supercurrent of a typical $\pi$-type vortex state is investigated at $T = 0$ K. We also study the critical temperature up to which a $\pi$-type vortex state can survive, as well as its distinct feature in the LDOS. The Anderson's approximation is applied to analyze the stability of the $\pi$-type vortex under finite magnetic field in the case less deep in the type-II limit. Moreover, the $k_F\xi_0$ versus $R$ phase diagram of the vortex type at $T=0$ K is investigated. Sec.~\ref{conclusion} gives our main conclusions.

\section{Theoretical Model}\label{theory}
Following Hayashi \textit{et al.}~\cite{Hayashi1998}, we consider an extreme
type-II superconducting nanocylinder, which allows us to ignore the vector
potential ${\bf A}({\bf r})$. Measuring the length scale (energy) by $\xi_0$
($\Delta_0$), we solve the following dimensionless BdG equations
\begin{equation}\label{bdg}
\left[
\begin{array}{cc}
\hat{H}_e & \Delta({\bf r}) \\
\Delta^*({\bf r}) & -\hat{H}^*_e
\end{array}
\right] \left[
\begin{array}{c}
u_\nu({\bf r}) \\
v_\nu({\bf r})
\end{array}
\right] = E_\nu
\left[
\begin{array}{c}
u_\nu({\bf r}) \\
v_\nu({\bf r})
\end{array}
\right],
\end{equation}
with
$\hat{H}_e = -\frac{1}{2k_F\xi_0}\nabla^2 - \mu_F$, $\Delta({\bf r})$ the
pair potential, $\mu_F=\frac{k_F\xi_0}{2}$ the Fermi energy, $E_\nu$ the
quasiparticle energy measured from $\mu_F$, $u_\nu(\bf r)$ and $v_\nu({\bf r})$ the electron- and
hole-like wave functions, respectively. Eqs. (\ref{bdg}) are calculated
self-consistently together with the pair potential equation
\begin{equation}\label{op}
 \Delta({\bf r}) = g\sum_{0<E_\nu<\hbar\omega_D} u_\nu({\bf r})
v^*_\nu({\bf r})\, {\rm tanh}\bigg(\frac{E_\nu}{2k_BT}\bigg),
\end{equation}
where $g$ is the coupling constant and $\omega_D$ the Debye frequency. In our numerical calculations, the Debye window around the Fermi level is set as $\hbar \omega_D = 10 \Delta_0$.

Due to the cylindrical symmetry of the sample, cylindrical coordinates, i.e.
$(\rho, \theta, z)$, are employed in our description. In this work we are
interested in an isolated single-quantum vortex, so, it is convenient to take the pair potential of the form
\begin{equation}\label{gap}
  \Delta({\bf r})=\Delta({\rho}){\rm e}^{-i\theta},
\end{equation}
where $\Delta(\rho)$ is the radial order parameter. Moreover, the quasiparticle wave functions are written as
\begin{eqnarray}
  u_\nu({\bf r}) &=\frac{1}{\sqrt{2\pi}} u_{jm}(\rho)e^{i(m-\frac{1}{2})\theta}e^{ik_zz}, \\
  v_\nu({\bf
r})&=\frac{1}{\sqrt{2\pi}}v_{jm}(\rho)e^{i(m+\frac{1}{2})\theta}e^{ik_zz}
\end{eqnarray}
with $\nu=\{j,m\}$, $j$ the radial quantum number, $2m$ an odd
integer~\cite{Caroli1964,Gygi1991}, and $k_z$ the wave number along the cylinder axis. The radial quasiparticle functions $u_{\nu}(\rho)$ and $v_{\nu}(\rho)$ are independent of $k_z$ by assuming a cylindrical Fermi surface~\cite{Gygi1991}. The boundary conditions are determined by the transverse quantum confinement which requires $u_{\nu}(R) = v_{\nu}(R) = 0$. So, $u_{\nu}(\rho)$ and $v_{\nu}(\rho)$ are expanded in terms of the Bessel functions as
\begin{eqnarray}
  u_{\nu}(\rho) &= \sum_{i}c_{\nu,i}\phi_{i,(m-\frac{1}{2})}(\rho) \\
  v_{\nu}(\rho) &= \sum_{i}d_{\nu,i}\phi_{i,(m+\frac{1}{2})}(\rho)\label{uv_exp}
\end{eqnarray}
with single-electron wave functions
$\phi_{i,n}(\rho)=\frac{\sqrt{2}}{RJ_{n+1}(\alpha_{i,n})}J_{n}(\frac{\alpha_{i,n
}\rho}{R})$, where $J_n(x)$ is the $n$th order Bessel function of the first kind,
$\alpha_{i,n}$ is its $i$th zero and $n=m\pm\frac{1}{2}$. Inserting Eq. (\ref{gap}) and the quasiparticle wave functions into the BdG Eqs. (\ref{bdg}), an eigenvalue problem of the coefficients
$c_{\nu,i}$ and $d_{\nu,i}$ is constructed, which is solved self-consistently by using Eq. (\ref{op}). Notice that there are three characteristic parameters: $k_F\xi_0$, $R$ and $T$, which control the extents of quantum limit, quantum confinement and thermal smearing, respectively.

\section{Results and discussions}\label{results}
\subsection{The $\pi$-type vortex at $T=0$ K}
%=================================================
\begin{figure}[t]
\centering
\includegraphics[scale=0.35]{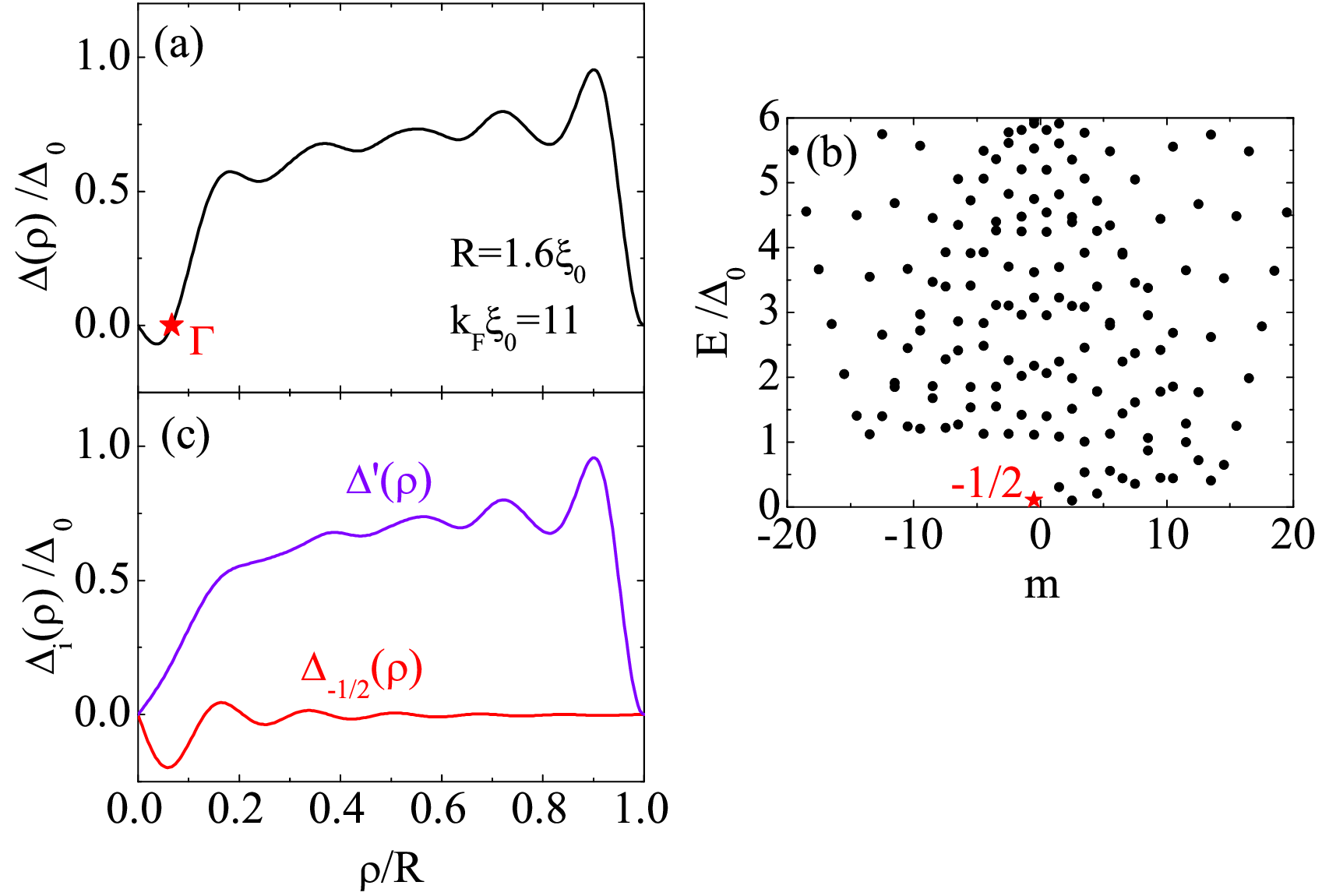}
\caption{(Color online) The $\pi$-type vortex state of $R=1.6\,\xi_0$ and $k_F\xi_0=11$: (a) the spatial radial order parameter $\Delta(\rho)$ and at $\Gamma$ point $\Delta(\rho)=0$; (b) the spatial supercurrent $j(\rho)$; (c) the quasiparticle energy spectrum with positive energy, in which the stared point is the chirality-breakdown bound state $m=-\frac{1}{2}$; (d) the radial order parameters contributed by the state $m=-\frac{1}{2}$ labeled as $\Delta_{-\frac{1}{2}}(\rho)$, and for all the other states labeled as $\Delta'(\rho)$.}
\label{fig1}
\end{figure}
%=================================================
% supercurrent $j(\rho)$ in (b) and quasiparticle energy spectrum $E_m$ in (c)
Fig. \ref{fig1} (a) shows the radial order parameter $\Delta(\rho)$ for a typical $\pi$-type vortex state of $k_F\xi_0=11$ and $R=1.6\,\xi_0$ at $T=0$ K. We see that $\Delta(\rho)$ becomes negative in the region from the center to the point $\Gamma$ at $\rho \approx0.066R$, while in the rest region it stays positive, i.e., $\Delta(\rho)$ clearly undergoes a $\pi$-phase shift around the core. The $\pi$-phase-shift area has a radius about half the Fermi wavelength $\frac{\lambda_F}{2} \sim \frac{1}{k_F}\approx0.057R$. This behavior is related to the breakdown of the chirality of the in-gap fermion state of azimuthal quantum number $m=-\frac{1}{2}$, which is labeled by the red star in the quasiparticle energy spectrum Fig. \ref{fig1}(b). The azimuthal quantum number ($m=-\frac{1}{2}$) and its quasiparticle energy ($E_{-\frac{1}{2}}=0.11\Delta_0$) have opposite signs. In fact, the state $m=-\frac{1}{2}$ shows up in Fig. \ref{fig1}(b), while the conjugated bound state $m=\frac{1}{2}$ disappears, i.e. its energy becomes negative. The chirality-breakdown bound state $m=-\frac{1}{2}$ with $E_{-\frac{1}{2}} > 0$ corresponds to the bound state $m=\frac{1}{2}$ with $E_{\nu} < 0$ since the BdG equations are invariant under the time-reversed transformation~\cite{Gygi1991}: $\{u({\bf r}), v({\bf r}), E_\nu\} \leftrightarrow  \{v^*({\bf r}), -u^*({\bf r}), -E_\nu\}$. Fig. \ref{fig1}(c) illustrates the contributions of the state $m=-\frac{1}{2}$ and all the other states to the whole $\Delta(\rho)$, labeled as $\Delta_{-\frac{1}{2}}(\rho)$ and $\Delta'(\rho)$, respectively. We see that $\Delta_{-\frac{1}{2}}(\rho=\frac{\lambda_F}{2})$ reaches $-0.20\Delta_0$, while $\Delta'(\frac{\lambda_F}{2})\approx0.17\Delta_0$.

\begin{figure}[t]
\centering
\includegraphics[scale=0.35]{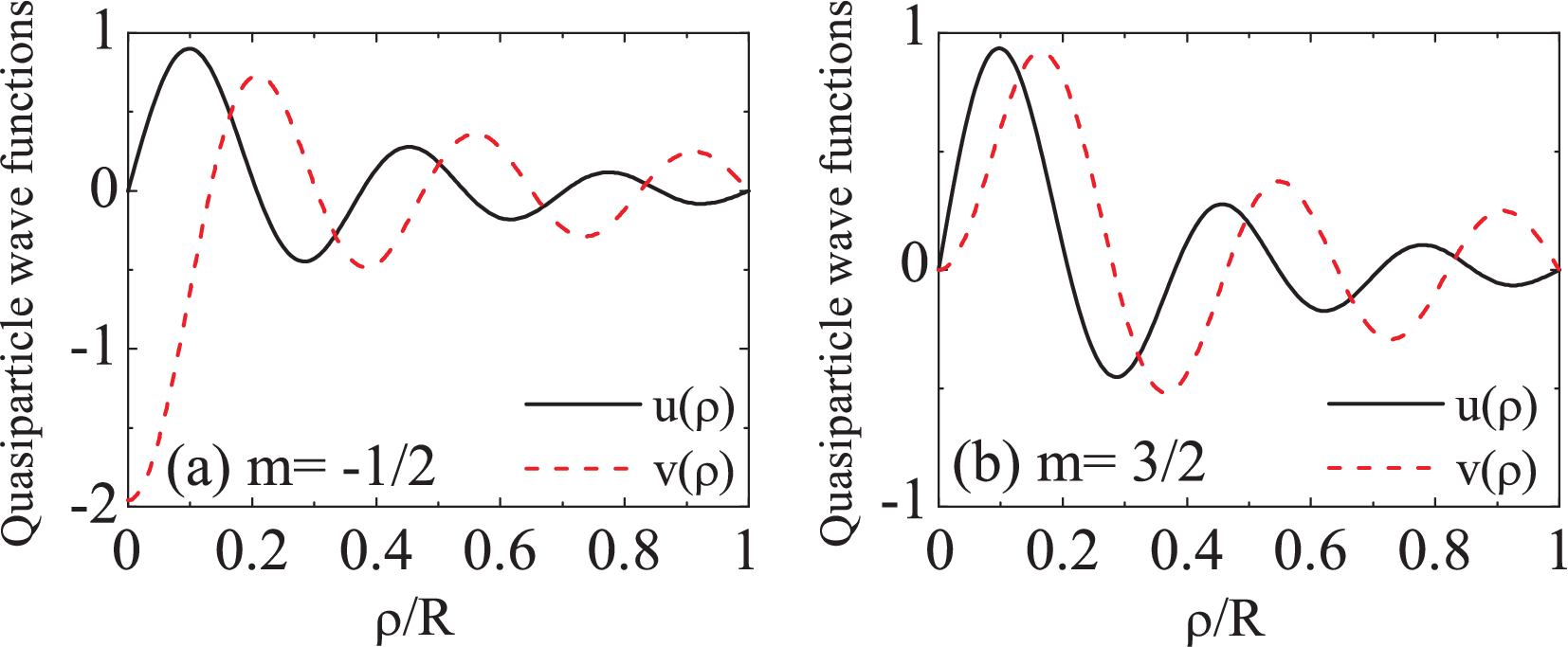}
\caption{(Color online) The spatial quasiparticle wave functions $u(\rho)$ and $v(\rho)$ of the chirality-breakdown bound state $m=-\frac{1}{2}$ (a) and normal bound state $m=\frac{3}{2}$ (b) for the $\pi$-type vortex state with $R=1.6\,\xi_0$ and $k_F\xi_0 = 11$. }
\label{fig2}
\end{figure}

In order to understand the negative contribution of the chirality-breakdown state $m=-\frac{1}{2}$ for $k_F\xi_0 = 11$ and $R = 1.6\xi_0$, its quasiparticle wavefunctions $u(\rho)$ and $v(\rho)$ are plotted in Fig. \ref{fig2}(a). We see that the $u(\rho)$ and $v(\rho)$ in the region $\rho<0.13\,R$ have opposite signs near the core, which results into negative value in $\Delta_{-\frac{1}{2}}(\rho)$ according to Eq.~(\ref{op}). However, the quasiparticle wave functions of all the other bound state near the core have the same sign, as well as the case for the state $m=\frac{3}{2}$ in Fig. \ref{fig2}(b).
%=================================================
%=================================================

%=================================================
\begin{figure}[b]
\centering
\includegraphics[scale=0.17]{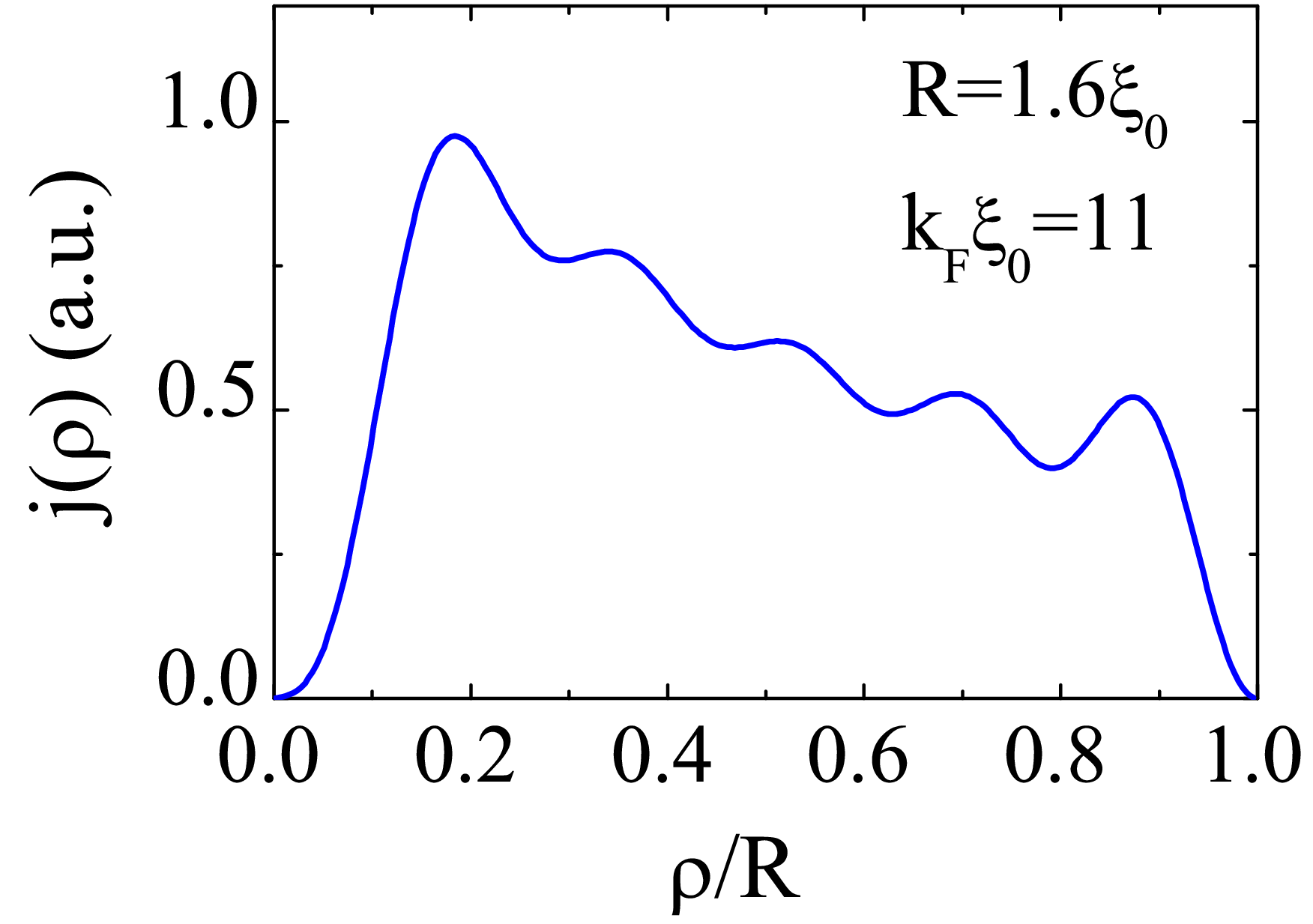}
\caption{(Color online) The spatial supercurrent $j(\rho)$ of the $\pi$-type vortex state of $R=1.6\,\xi_0$ and $k_F\xi_0=11$.}
\label{fig3}
\end{figure}
%=================================================
On the other hand, as seen from Fig.~\ref{fig3}(b), the supercurrent in the $\pi$-phase-shift region shows a $\rho^3$ behavior, rather than the conventional linear behavior of a macroscopic single-quantum vortex state. This extraordinary characteristic in $j(\rho)$ is also caused by the chirality-breakdown state of $m=-\frac{1}{2}$. To explore the details, we start with the contribution of a single fermion state ${\nu}$ to $j(\rho)$, which is of the form
\begin{equation}\label{curj}
 j_\nu(\rho) \propto \frac{1}{\rho}\bigg[
(m-\frac{1}{2})u_{\nu}^2f_{\nu}-(m+\frac{1}{2})v_{\nu}^2\big(1-f_{\nu}\big)
\bigg]
\end{equation}
with $f_{\nu}=f(E_\nu, T)$ the Fermi distribution function. For all states with positive quasiparticle energy $f_{\nu}=0$ at $T=0$ K, only the second term in Eq. (\ref{curj}) is non-zero. So, the chirality-breakdown bound state $m=-\frac{1}{2}$ gives no contribution to $j(\rho)$ at $T=0$ K according to Eq. (\ref{curj}). Near the vortex core, the major contribution to $j(\rho)$ comes from the in-gap fermion $m=\frac{3}{2}$, whose hole-like wave function $v_{\frac{3}{2}}(\rho)$ is a combination of the $2$nd-order Bessel functions of the first kind $J_2(\frac{\alpha_{i,2}}{R}\rho)$. As $x\rightarrow0$, $J_2(x)\sim x^2$. By inserting these above relations into Eq. (\ref{curj}), we obtain a $\rho^3$ behavior of $j(\rho)$ close to the vortex core.

\subsection{The $\pi$-type vortex at finite temperatures}
When switching on thermal smearing, to what extent the $\pi$-type vortex state can survive is of great interest and importance. As increasing $T$, the first term in Eq. (\ref{curj}) for the chirality-breakdown bound state $m=-\frac{1}{2}$ turns to be nonzero. Thus, the linear dependence of $j(\rho)$ gradually shows up around the core, and finally takes the major trend at a certain temperature $T_\tau$, i.e. the characteristic of the supercurrent of $\pi$-type vortex state gets smeared out. The value of $T_\tau$ can be set by the temperature at which the state $m=-\frac{1}{2}$ give half contribution of the electron-like part, i.e. $f(E_{-\frac{1}{2}},T_\tau)=\frac{1}{2}$. Such criteria leads to $T_\tau = E_{-\frac{1}{2}}/k_B$. For instance, for the $\pi$-type vortex state of $k_F\xi_0=11$ and $R=1.6\,\xi_0$, $T_{\tau} = 0.19\,T_c$.
%=================================================
\begin{figure}[t]
\centering
\includegraphics[scale=0.3]{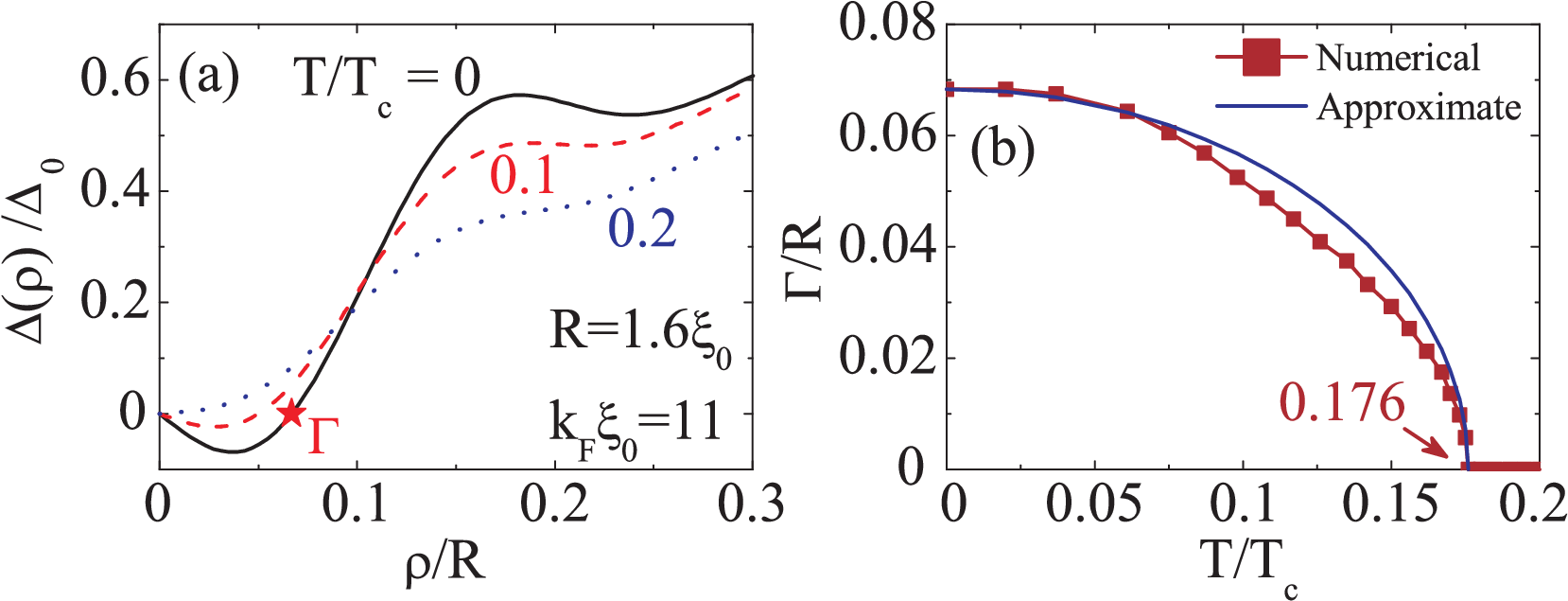}
\caption{(Color online) (a) Spatial radial order parameters of the $\pi$-type vortex state with $R = 1.6\,\xi_0$ and $k_F\xi_0 = 11$ at $T/T_c = 0$, $0.1$ and $0.2$. (b) Temperature dependence of numerically extracted $\Gamma$ data (solid symbolled line), the blue curve is an analytical function $\Gamma(T)=\Gamma(0)\sqrt{1-\big(\frac{T}{T_\tau'}\big)^2}$ with $T_\tau'=0.18\,T_c$.}
\label{fig4}
\end{figure}
%=================================================
This criterion is quantitatively supported by numerical results. The temperature dependence of $\Gamma$ value is shown in Fig.~\ref{fig1}(a), which goes to zero when the $\pi$-type vortex state is terminated. Fig. \ref{fig4}(a) shows the zoom of $\Delta(\rho)$ in the core region at $T/T_c = 0,\, 0.1$ and $0.2$ for $k_F\xi_0 = 11$ and $R=1.6\,\xi_0$. We see that the $\pi$-phase-shift region of the radial order parameter decreases as $T$ increases, and it finally disappears at a critical temperature. More detail about the temperature dependence of the $\pi$-phase-shift region, i.e. $\Gamma(T)$, is shown as the squared line in Fig. \ref{fig4}(b). As seen from in Fig. \ref{fig4}(b), the numerical results agree very well with the curve $\Gamma(T)=\Gamma(0)\sqrt{1-\big(\frac{T}{T_\tau'}\big)^2}$ with $\Gamma(0) = 0.066R$ and $T_\tau' = 0.18\,T_c$. Here, $T_\tau'$ is the exact value of the maximum temperature for the survival of the $\pi$-type vortex state, and matches the analytical $T_\tau$ very well, which means that $T_\tau$ is a very good approximation.

\subsection{The LDOS of the $\pi$-type vortex}
\begin{figure}[t]
\centering
\includegraphics[scale=0.35]{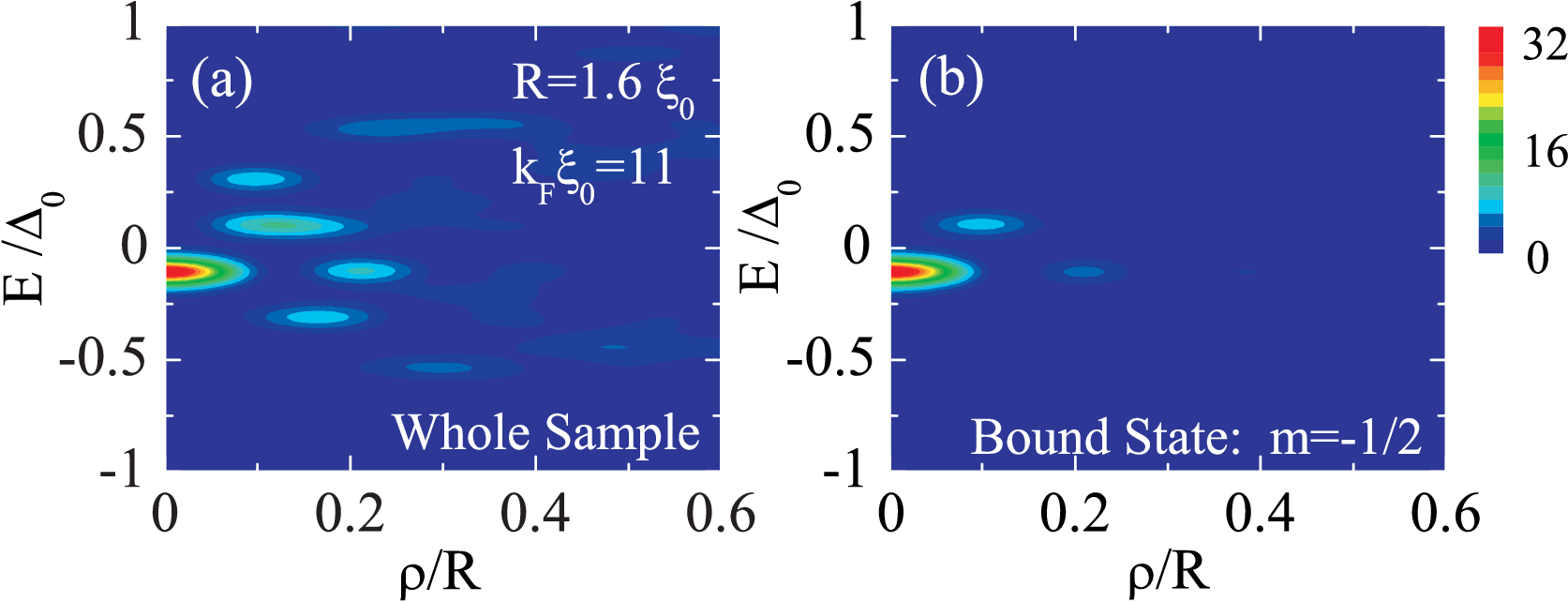}
\caption{(Color online) The LDOS spectra for the $\pi$-type vortex state with $R = 1.6 \xi_0$ and $k_F\xi_0 = 11$ at $T=0.05T_c$: (a) the whole sample; (b) the single chirality-breakdown bound state $m=-\frac{1}{2}$.}
\label{fig5}
\end{figure}

To find the evidence of the chirality-breakdown bound state or even $\pi$-type vortex state in experiments, Fig. \ref{fig5} (a) and (b) plot the LDOS spectra of the whole $\pi$-type vortex state and the single
chirality-breakdown bound state $m=-\frac{1}{2}$ for $R = 1.6\,\xi_0$ and $k_F\xi_0 = 11$ at $T=0.05T_c$, respectively.
%=================================================
%=================================================
The LDOS for the whole sample in Fig. \ref{fig5}(a) shows similar particle-hole asymmetry and isolated peaks as in quantum limit in the absence of quantum confinement~\cite{Hayashi1998}, but there is a significant difference. Here, the major peak at the vortex center $\rho = 0$ locates at $E\simeq -E_{-\frac{1}{2}}=-0.11\Delta_0$, i.e. on the positive energy side, while in Fig. 5 of Ref. \cite{Hayashi1998} the largest peak associated with the bound state $m=\frac{1}{2}$ is on the negative energy side. Moreover, as seen from Fig. \ref{fig5}(b), the major peak at $\rho = 0$ and $E = -0.11\Delta_0$ is mainly contributed by the chirality-breakdown bound state $m=-\frac{1}{2}$. This negative-shifted zero-bias peak feature qualitatively differs from the previous studies, and can be checked by STM experiments.

\subsection{The phase diagram of the vortex type}\label{phase}
The chirality-breakdown bound state $m=-\frac{1}{2}$ plays a crucial role in the appearance of the $\pi$-type vortex state with $k_F\xi_0 = 11$ and $R = 1.6\xi_0$. But, what controls the appearance of the chirality-breakdown bound state? To answer this question, we can rely on the analysis similar to the Anderson approximation~\cite{Anderson1959}, which reduces Eqs. (\ref{uv_exp}) into a single-component expression. This approximation is supported well by our numerical calculation: at $T=0$ K for the chirality-breakdown bound state $m=-\frac{1}{2}$ in the $\pi$-type vortex state with $R = 1.6\xi_0$ and $k_F\xi_0 = 11$ there is only one major component in the expansion of $u(\rho)$ and $v(\rho)$ in Eqs. (\ref{uv_exp}). The single-electron state $\phi_{\,6,0}(\rho)$ with $|c|^2=50.0\%$ and $\phi_{\,6,1}(\rho)$ with $|d|^2=43.5\%$ together give a possibility of $93.5\%$ to the bound state $m=-\frac{1}{2}$. By inserting this approximation into the BdG Eqs. (\ref{bdg}) and the self-consistent condition Eq. (\ref{op}), the physical quasiparticle energy is obtained analytically
\begin{equation}\label{uv_and}
  E_{\nu} = \frac{\epsilon^-_{\nu} - \epsilon^+_{\nu}}{2} + \sqrt{\bigg(\frac{\epsilon^-_{\nu} +
\epsilon^+_{\nu}}{2}\bigg)^2 + \bar{\Delta}_{\nu}^2},
\end{equation}
where $\epsilon^\pm_{\nu}=\frac{\alpha^2_{j,(m\pm \frac{1}{2})}}{2k_F\xi_0 R^2}-\mu_F$ are the energies of two neighboring single-electron states and $\bar{\Delta}_{\nu} = \int_0^R  \rho\, d\rho \,\phi^*_{j,(m-\frac{1}{2})}(\rho)
\Delta(\rho) \phi_{j,(m+\frac{1}{2})}(\rho)$. As seen from Eq. (\ref{uv_and}), the competition between $\bar{\Delta}_{\nu}$ and the quantization of the single-electron energy ($\epsilon^\pm_{\nu}$) induced by quantum confinement determines the sign of quasiparticle energy $E_{\nu}$ of the state, i.e. the effect of destruction or construction to the superconductivity~\cite{Shanenko2008}. Particularly, the chirality-breakdown bound state $m=-\frac{1}{2}$ turns up, i.e. the energy of its conjugated state $m=\frac{1}{2}$ becomes negative, when $\bar{\Delta}_{\nu}^2 <-\epsilon^-_{j,m=\frac{1}{2}}\times \epsilon^+_{j,m=\frac{1}{2}}$. It implies that $\mu_F$ ($=k_F\xi_0/2$) should locate between the two bands with energies $\epsilon^-_{j,m=\frac{1}{2}}$ and $\epsilon^+_{j,m=\frac{1}{2}}$. We also see that the energy interval between $\epsilon^-_{j,m=\frac{1}{2}}$ and $\epsilon^+_{j,m=\frac{1}{2}}$ depends on $R$. Therefore, the appearance of the chirality-breakdown bound state $m=-\frac{1}{2}$ and the $\pi$-type vortex state is controlled by $k_F\xi_0$ and $R$.

Now, we study the question: when will a single-quantum $\pi$-type vortex state appear in a nanoscale extreme type-II $s$-wave superconductor? The phase diagram of the vortex type as $R$ versus $k_F\xi_0$ at $T=0$ K is shown in Fig. \ref{fig5}. We find that numerous $\pi$-type vortex state (red/dark) area discretely distributes mainly in the region of $R < 2\,\xi_0$ and $k_F\xi_0$ up to $50$.
\begin{figure}[t]
\centering
\includegraphics[scale=0.22]{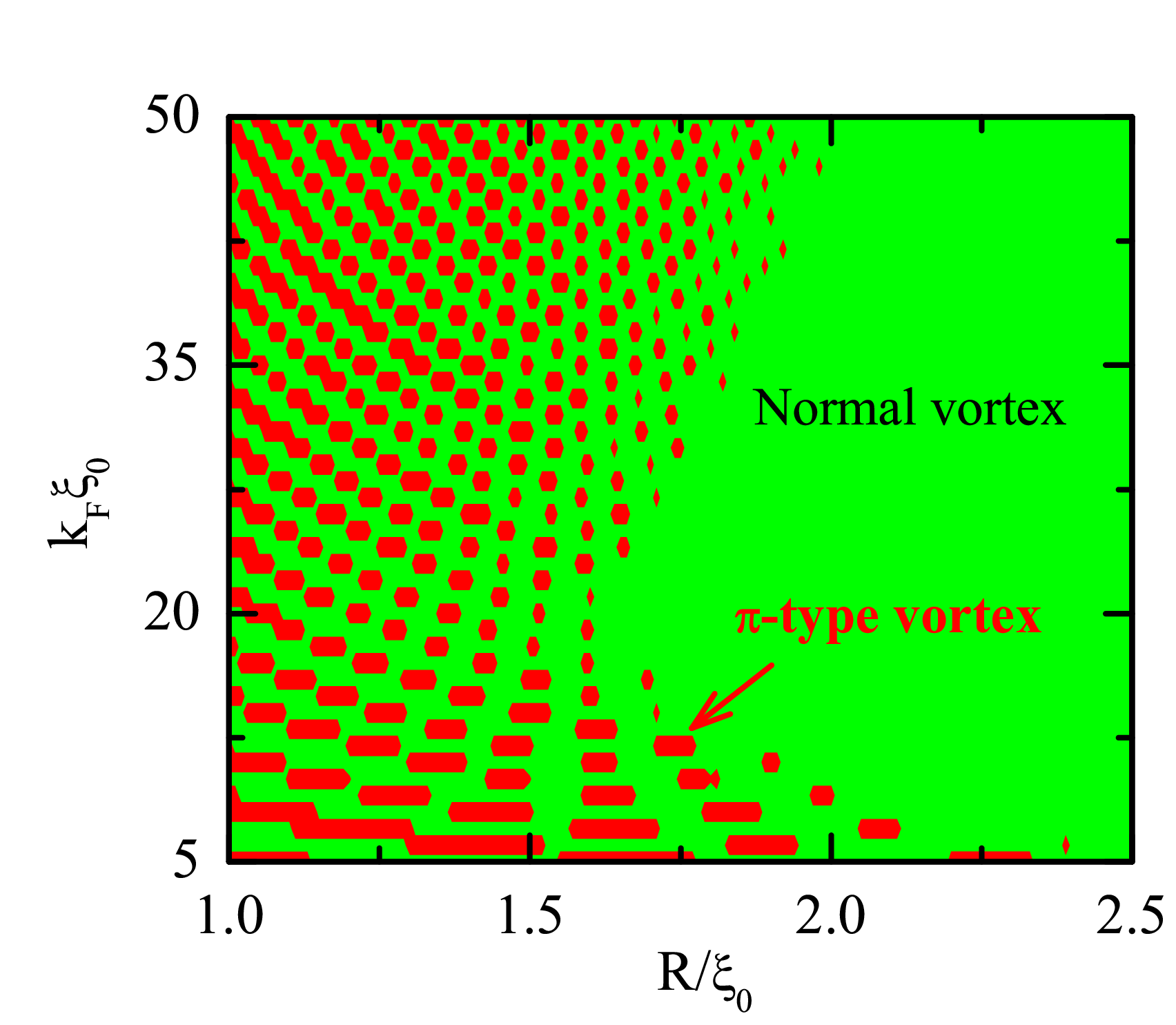}
\caption{(Color online) Phase diagram of the vortex type as $k_F\xi_0$ versus $R$ at $T = 0$ K. The red (dark) region represents the $\pi$-type vortex state region, while the rest green (grey) region indicates the normal vortex region.}
\label{fig6}
\end{figure}
Such discrete structure is caused by the quantization of the single-electron energy spectrum. As decreasing $R$ for a fixed $k_F\xi_0$, the area of the $\pi$-type vortex state increases because of the enhancement energy gap between quantized single-electron states (see Eq.~(\ref{uv_and})). When $R$ increases and reaches a certain value $R_{max}$, the $\pi$-type vortex state vanishes. $R_{max}$ varies with $k_F\xi_0$ as a result of competition between quantum confinement and $\bar{\Delta}_{\nu}$ in Eq.~(\ref{uv_and}). For $k_F\xi_0 \sim 50$, the single-electron energy spectrum is almost continuous, so, it is difficult to find the chirality-breakdown in-gap state $m=-\frac{1}{2}$ according to Eq.~(\ref{uv_and}), i.e. the $\pi$-type vortex state. But, one may steadily observe the $\pi$-type vortex state at a proper range of $R$ for a clean $s$-wave extreme type-II superconductor with relatively low $k_F\xi_0$, e.g. a typical $A$-$15$ compound V$_3$Si~\cite{Corcoran1994} with $k_F\xi_0\simeq12$, or even YBCO~\cite{Maggio-Aprile1995} with $k_F\xi_0 \sim 4$. The $\pi$-type vortex state also can be found in the high $k_F\xi_0$ region such as for Pb nanoislands~\cite{Cren2009,Cren2011}, however, it requires well-controlled boundary roughness according to the narrow area in Fig.~\ref{fig6}.

\subsection{The $\pi$-type vortex in less deep type-II regime}
Next, we consider the stability of the $\pi$-type vortex in less deep type-II regime. Its exact solution requires to solve the full BdG equations self-consistently including the magnetic field. However, the calculation of such procedure is very heavy and time-consuming. For simplicity, we employ the Anderson's approximation as in Sec. \ref{phase}, and assume that the magnetic field participates but varies slowly in space in the case of less deep type-II regime. So, the vector potential is approximately taken as ${\bf A}(\rho) = \frac{H_0\rho}{2}{\bf e}_{\theta}$ with $H_0$ the applied field and ${\bf A}(\rho)$ measured in units of $\frac{\Phi_0}{2\pi\xi_0^2}$ and ${\bf A}(\rho)$ and $\frac{\Phi_0}{2\pi\xi_0}$, respectively, where $\Phi_0=\frac{ch}{2e}$ is the single quantum flux. The single-electron Hamiltonian in the BdG Eqs. (\ref{bdg}) involving the field reads
\begin{equation}\label{Hmag}
  \hat{H}_e = -\frac{1}{2k_F\xi_0} \nabla^2 + i\frac{1}{2k_F\xi_0}{\bf A}\cdot \nabla + \frac{1}{8k_F\xi_0}{\bf A}^2.
\end{equation}
We only consider the case with sufficiently low $H_0$, thus, it is reasonable to neglect the third term in the above equation. Insert Eq. (\ref{Hmag}) and the following approximate quasiparticle wavefunctions
\begin{eqnarray}
  u_\nu ({\bf r}) &= \frac{c_i}{\sqrt{2\pi}}\psi_{i,(m-\frac{1}{2})(\rho)}e^{i(m-\frac{1}{2})\theta}e^{ik_zz} \\
  v_\nu ({\bf r}) &= \frac{d_i}{\sqrt{2\pi}}\psi_{i,(m+\frac{1}{2})}(\rho)e^{i(m+\frac{1}{2})\theta}e^{ik_zz}
\end{eqnarray}
into the BdG Eqs. (\ref{bdg}) and the self-consistent condition Eq. (\ref{op}), and we obtain the physical quasiparticle energy analytically
\begin{align}\label{Emag}
  E_{\nu} = &\frac{\epsilon^-_{\nu} - \epsilon^+_{\nu}}{2} - \frac{mH_0}{4k_F\xi_0} \\ \nonumber
  &+ \sqrt{\bigg(\frac{\epsilon^-_{\nu} +
\epsilon^+_{\nu}}{2}+\frac{H_0}{8k_F\xi_0}\bigg)^2 + \bar{\Delta}_{\nu}^2} ,
\end{align}
where $\epsilon^\pm_{\nu}$ and $\bar{\Delta}_{\nu}$ defined in Eq. (\ref{uv_and}) are independent of $H_0$. Such expression is complicated for analyzing the influence of the applied field on $E_{\nu}$. Thus, we take a crude but efficient approximation
\begin{align}\label{Emag2}
  E_{\nu} \sim &\frac{\epsilon^-_{\nu} - \epsilon^+_{\nu}}{2} - \frac{(2m-1)H_0}{8k_F\xi_0} \\ \nonumber
  &+ \sqrt{\bigg(\frac{\epsilon^-_{\nu} +
\epsilon^+_{\nu}}{2}\bigg)^2 + \bar{\Delta}_{\nu}^2}.
\end{align}
When the vortex state is $\pi$-type at zero field, i.e. only the in-gap state $m=\frac{1}{2}$ has negative energy, it keeps $\pi$-type as $H_0$ increases until the energies of other states become negative. Assuming the minimum energy of the in-gap fermion is $E_{m}$, then, the minimal field killing the $\pi$-type is $H_{0,min} = \frac{8E_{m}}{2m-1}k_F\xi_0 \frac{\Phi_0}{2\pi\xi_0^2}$. For example, taking $E_{m=\frac{3}{2}}\sim0.1$ and $k_F\xi_0\sim10$, one obtains $H_{0,min} \sim 4\,\frac{\Phi_0}{2\pi\xi_0^2}$. It implies that if the radius takes $R=\xi_0$, the flux passing through the sample is $2\Phi_0$. Therefore, for these parameters, the $\pi$-type vortex may remain stable under a field in the range from $0$ to $4\frac{\Phi_0}{2\pi\xi_0^2}$.

\section{Conclusions}\label{conclusion}
To conclude, by numerically solving the Bogoliubov-de Gennes equations, we demonstrated that a single-quantum vortex state in a nanoscale cylindrical extreme type-II $s$-wave superconductor may undergo a $\pi$-phase shift in the radial order parameter near the core. We refer to it as $\pi$-type vortex state induced by quantum-size effect. The chirality-breakdown bound state of azimuthal quantum number $m=-\frac{1}{2}$ plays the key role on such behavior. Such state shows two distinctive features: first, the supercurrent has a $\rho^3$ behavior near the core, rather than the conventional linear dependence; second, the LDOS exhibits a significant negative-shifted zero-bias peak. The $\pi$-type vortex state may survive thermal smearing when $T/T_c < E_{-\frac{1}{2}}/\Delta_0$ with $E_{-\frac{1}{2}}$ quasiparticle energy of bound state $m=-\frac{1}{2}$, and it may remain stable under sufficiently week magnetic field in the case less deep in the type-II limit. Based on the numerically calculated phase diagram of vortex type as $k_F\xi_0$ versus $R$, we may expect to find $\pi$-type vortex state in extreme type-II superconductors and superfluids with low $k_F\xi_0$ ($\sim10$) and $R < 2\,\xi_0$, e.g. Pb nanoislands~\cite{Cren2009,Cren2011}, ultracold superfluid Fermi gases in cigar-shaped and pancake-shaped atomic traps~\cite{Bloch2008}, or even high-$T_c$ superconductors, e.g. YBCO~\cite{Maggio-Aprile1995}.

This work was supported by the National Natural Science Foundation of China under Grant No. NSFC-11304134, and Zhejiang Provincial Natural Science Foundation(No. Y14F030005). Y. Chen acknowledges the valuable discussion with Prof. Dr. Arkady Shanenko and Prof. Dr. Shi-ping Zhou.


\section*{References}

\begin{thebibliography}{99}

\bibitem{Ryazanov2001} V.~V. Ryazanov, V.~A. Oboznov, A.~Y. Rusanov, A.~V. Veretennikov, A.~A. Golubov, and J.~Aarts, Phys. Rev. Lett. {\bf 86}, 2427 (2001).%1

\bibitem{Krivoruchko2002} V.~N. Krivoruchko and E.~A. Koshina, Phys. Rev. B {\bf 66}, 014521 (2002). %2

\bibitem{Buzdin2005} A.~I. Buzdin, Rev. Mod. Phys. {\bf 77}, 935 (2005). %3

\bibitem{VanHarlingen1995} D.~J. Van~Harlingen, Rev. Mod. Phys. {\bf 67}, 515 (1995). %4

\bibitem{Balatsky2006} A.~V. Balatsky, I.~Vekhter, and J.-X. Zhu, Rev. Mod. Phys. {\bf 78}, 373 (2006). %5

\bibitem{Caroli1964} C.~Caroli, P.~D. Gennes, and J.~Matricon, Physics Letters {\bf 9}, 307 (1964). %6

\bibitem{Hess1989} H.~F. Hess, R.~B. Robinson, R.~C. Dynes, J.~M. Valles, and J.~V. Waszczak, Phys. Rev. Lett. {\bf 62}, 214 (1989).%7

\bibitem{Hess1990} H.~F. Hess, R.~B. Robinson, and J.~V. Waszczak, Phys. Rev. Lett. {\bf 64}, 2711 (1990).%8

\bibitem{Roditchev2015} D. Roditchev, C. Brun, L. Serrier-Garcia, J. C. Cuevas, V. H. L. Bessa, M. V. Milo$\breve{s}$evi$\acute{c}$, F. Debontridder, V. Stolyarov and T. Cren, Nature Physics {\bf 11}, 332 (2015).

\bibitem{Kramer1974} L.~Kramer and W.~Pesch, Z. Phys. {\bf 269}, 59 (1974). %9

\bibitem{Volovik1993} G.~L. Volovik, JETP Lett. {\bf 58}, 455 (1993). %10

\bibitem{Nygaard2003} N.~Nygaard, G.~M. Bruun, C.~W. Clark, and D.~L. Feder, Phys. Rev. Lett. {\bf 90}, 210402 (2003). %11

\bibitem{DeBlasio1999} F.~V. De~Blasio and O.~Elgar\o{}y, Phys. Rev. Lett. {\bf 82}, 1815 (1999).%12


\bibitem{Guo2004} Y.~Guo, Y.-F. Zhang, X.-Y. Bao, T.-Z. Han, Z.~Tang, L.-X. Zhang, X.-G. Zhu, E.~G. Wang, Q.~Niu, Z.~Q. Qiu, et~al., Science {\bf 306}, 1915 (2004).%13

\bibitem{Shanenko2007} A.~A. Shanenko, M.~D. Croitoru, R.~G. Mints, and
  F.~M. Peeters, Phys. Rev. Lett. {\bf 99}, 067007 (2007). %14

\bibitem{Chen2009} Y.~Chen, M.~D. Croitoru, A.~A. Shanenko, and F.~M. Peeters, J. Phys.: Condens. Matter {\bf 21}, 435701 (2009). %15

\bibitem{Chen2010} Y.~Chen, A.~A. Shanenko, and F.~M. Peeters, Phys. Rev. B {\bf 81}, 134523 (2010). %16

\bibitem{Cren2009} T.~Cren, D.~Fokin, F.~Debontridder, V.~Dubost, and D.~Roditchev, Phys. Rev. Lett. {\bf 102}, 127005 (2009). %17

\bibitem{Cren2011} T.~Cren, L.~Serrier-Garcia, F.~Debontridder, and D.~Roditchev, Phys. Rev. Lett. {\bf 107}, 097202 (2011). %18

\bibitem{Zhang2012} L.-F. Zhang, L. Covaci, M. V. Milo$\breve{s}$evi$\acute{c}$, G. R. Berdiyorov, and F. M. Peeters, Phys. Rev. Lett. 109, 107001 (2012).

\bibitem{Zhang2013} L.-F. Zhang, L. Covaci, M. V. Milo$\breve{s}$evi$\acute{c}$, G. R. Berdiyorov, and F. M. Peeters, Phys. Rev. B 88, 144501 (2013).

\bibitem{Chen2014} Y.~Chen, A.~A. Shanenko, and F.~M. Peeters, Phys. Rev. B {\bf 89}, 054513 (2014). %19

\bibitem{Chen2015} Y.~Chen, W.~Hong-Yu, F.~M. Peeters, and A.~A. Shanenko, Journal of Physics: Condensed Matter {\bf 27}, 125701 (2015). %20

\bibitem{Bloch2008} I.~Bloch, J.~Dalibard, and W.~Zwerger, Rev. Mod. Phys. {\bf 80}, 885 (2008).


\bibitem{Hayashi1998} N.~Hayashi, T.~Isoshima, M.~Ichioka, and K.~Machida, Phys. Rev. Lett. {\bf 80}, 2921 (1998).

\bibitem{Gygi1991} F.~Gygi and M.~Schl\"uter, Phys. Rev. B {\bf 43}, 7609 (1991).

\bibitem{Anderson1959} P.~W. Anderson, J. Phys. Chem. Solids {\bf 11}, 26 (1959).

\bibitem{Shanenko2008} A.~A. Shanenko, M.~D. Croitoru, and F.~M. Peeters, Phys. Rev. B {\bf 78}, 024505 (2008).

\bibitem{Corcoran1994} R.~Corcoran, N.~Harrison, S.~M. Hayden, P.~Meeson, M.~Springford, and P.~J. van~der Wel, Phys. Rev. Lett. {\bf 72}, 701 (1994).

\bibitem{Maggio-Aprile1995} I.~Maggio-Aprile, C.~Renner, A.~Erb, E.~Walker, and O.~Fischer, Phys. Rev. Lett. {\bf 75}, 2754 (1995).

\end{thebibliography}
\end{document}